\def\apj{ApJ\ }
\def\apjl{ApJL\ }  

\def\aa{A\&A\ }   
\def\mn{MNRAS } 
\def\x{{\bf x}}

\documentstyle{l-aa}
\input psfig

\begin{document}
 
\thesaurus{02 (11.17.3; 12.04.1; 12.12.1) }      
\title{ Abundance and Clustering of QSOs in Cosmic Structure
Formation Models}
%
%\subtitle{}
\author{ Hong-Guang Bi\inst{1} and Li-Zhi Fang\inst{2}} 
\institute{ $^{1}$ Center for Astrophysical Sciences, Johns Hopkins
 University, Maryland, MD 21218, USA, and Beijing Astronomical 
         Observatory, Beijing 100080, China\\
            $^{2}$  Department of Physics, University of Arizona, Tucson,
           AZ 85721, USA} 
\offprints{L.Z.Fang}
\date{Received ... 1997 / Accepted ... 1997} 
\maketitle
\markboth{Bi and Fang: Abundance and Clustering of QSOs}{}
\begin{abstract} 
When combining the observations of spatial abundance and two-point correlation
function of QSOs, we can effectively set constraints on models of cosmic 
structure formation. Both the abundance of gravitationally confined halos
and their two-point correlation functions can be calculated in the
conventional Press-Schechter formalism. We apply this method to examine the
properties of possible host halos of QSOs in three popular models: the
standard cold dark
matter (SCDM) model, the low density flat cold dark matter (LCDM) model and
the cold-plus-hot dark matter (CHDM) model. The LCDM and CHDM models are
normalized to the COBE-DMR observations, and the SCDM is normalized to
$\sigma = 0.58$. We find that the SCDM and LCDM models can pass the
abundance-plus-correlation test for QSOs. However, the CHDM are difficult
to produce host halos to fit with the number density of high redshift
QSOs and their clustering on large scales (10 h$^{-1}$ Mpc) simultaneously.
We studied various mechanisms, originated both gravitationally and
non-gravitationally, which may lead to a biasing of the halo clustering. We
conclude that these effects are too weak in order to release the
trouble of the CHDM models.

\end{abstract}

\maketitle

\section {Introduction}

Recently, we have developed a method to test popular dark-matter models of
structure formation using the number density and the two-point correlation 
function of high redshift objects (Bi \& Fang 1996). We found an
approximate expression of two-point correlation function of mass and 
collapsed halos in the Press-Schechter formalism (Press \&
Schechter 1974, hereafter PS). In this approximation, the nonlinear
gravitational interaction was treated as the sum of various individual
spherical top-hat clustering. These top-hat spheres consist of both collapsed 
PS halos and uncollapsed regions. Moreover, the bias that massive
PS halos have stronger correlation than the background mass can naturally be
introduced by considering that no collapsed halo of mass $M$ exists in
initial regions (or top-hat spheres) of mass less than $M$.

This method was applied to CIV absorption systems in QSO spectra.
Because CIV systems should be hosted by collapsed halos, one can obtain an 
lower limit to the spatial number density of these host halos from the
observed CIV number density. This requires that the hosts of CIV should
consist of halos with mass as low as $M_{th}$. On the other hand, the two-point
correlation of halos is stronger when the mass of the halos is larger.
The observed two-point correlation functions set a lower limit, $M_{co}$,
to the mass of host halos. Obviously, a reasonable model has to give
$M_{th} > M_{co}$. The standard cold dark matter (SCDM) model
normalized to $\sigma = 0.58$ and the low density flat cold dark matter
(LCDM) model can pass this test. However, for the cold-plus-hot dark matter
(CHDM) models with
parameters $\Omega_c = 0.7$ and $\Omega_b = 0.3$, or $\Omega_c = 0.8$ and
$\Omega_b = 0.2$, the two-point correlation functions of halos with mass
$M_{th}$ are too small to explain the observed correlation
functions. In order to have enough number of collapsed halos to host CIV
systems, $M_{th}$ should not be larger than $10^{11}\ M_{\odot}$. But
the observed two-point correlation function on the scales of
$\Delta v \sim 300 - 1,000$ km/s indicates that $M_{co}$ should
not be less than $10^{12}\ M_{\odot}$.

In this paper, we study the same problem, but using QSOs as the discriminator.
The topic of finding constraints on dark matter models from QSOs is not new. 
As early as 1980s, the highest redshift of QSOs was used to rule out
the model of hot dark matter (HDM), because the HDM predicted that  collapsed
halos cannot form at redshift larger than 3, which is much less than
the redshifts of many existing QSOs. Consequently, the formation of high 
redshift QSOs cannot be explained in the HDM model (e.g. Efstathiou \& Rees 
1988). This redshift test is passed for current dark matter models.
For instance, the tilted CDM, scaled CDM and CHDM models all are able to 
produce with the observed abundance of QSOs (Nusser \& Silk 1993).
However, the QSO abundance test alone is not free from the uncertainties of
discriminating among the models like SCDM, LCDM and CHDM. While some works
claimed that the CHDM model is consistent with the QSO abundance, other
works found oppositely (Ma \& Bertschinger 1994). We will show that adding a
clustering test to the abundance fitting will reduce the number of free
parameters in the discrimination, and gives much better results.

\subsection{ Abundance of QSOs}

The spatial number density of collapsed halos can be calculated from
the standard PS theory. We define $\delta (\x, z)$ to be the 3-D density
fluctuation field of dark matter extrapolated to redshift $z$ assuming linear
evolution. A density field $\delta_R(\x, z)$, representing the smoothed
fluctuation on scale $R$, can be derived from $\delta (\x, z)$ by
%eq1
\begin{equation}
\delta_R (\x, z) = \frac{1}{V_R} \int \delta (\x', z) W(R; \x' -\x ) d\x',
\end{equation}
where the function $W(R; \x_1 -\x)$ is the top-hat window for the comoving 
volume $V_R = 4\pi R^3/3$. The total mass within $V_R$ on average is
$M = V_R \rho_0$, where $\rho_0$ is the mean density at the present
if the scaling factor of the universe is set to be unity at $z=0$.

For Gaussian perturbations, the mass fluctuation within a top-hat window of
radius $R$ is described by variance $\sigma^2(R,z)$, which is determined by
the initial density spectrum $P(k)$ and normalization factor $\sigma_8$, i.e.
$\sigma (R= 8$ h$^{-1}$Mpc). In the case of $\Omega =1$ Einstein-de Sitter
universe, the linear evolution of the variance is
$\sigma^2(R,z) \propto (1+z)^{-2}$. Thus, the fraction of the total
mass $\rho_0 \Delta \x$ having fluctuations larger than a given $\delta _c$
in an arbitrary spatial domain $\Delta \x $ is
%eq2
\begin{equation}
F_R = \int ^\infty_{\delta_c(z)} \frac{1}{\sqrt{2\pi} \sigma(R,z)}
      \exp{\left(-\frac{\delta_R ^2}{2\sigma^2(R,z)}\right)} d\delta_R.
\end{equation}

\begin{table*}
%\begin{center}
\caption{ Parameters of the cosmological models}
%\begin{flushleft}
\begin{tabular}{lccc}
\hline
        & SCDM   & LCDM  &  CHDM \\
\hline
$h$     &  0.5   &  0.75 &   0.5  \\
$\Omega_0$ & 1 & 0.3 &1 ($\Omega_\nu=0.3$)\\
$\lambda_0$ & 0 & 0.7 & 0  \\
$\sigma_8$ & 0.58 & 1 & 0.55  \\
\hline
\end{tabular}
%\end{flushleft}
%\end{center}
\end{table*}

If we take $\delta_c(z)$ to be the critical overdensity for the
collapse of the spherical mass $M$ at $z$, $F_R \cdot \rho_0 \Delta \x$ 
should be identified as the sum of masses of all collapsed halos, each of 
which is massive greater than $M$. For the Einstein-de Sitter universe, we 
have $\delta_c(z)=(1+z) 1.69$. For the flat $\Lambda$ universe, function 
$\delta_c(z)$ is calculated by Bi \& Fang (1996). The differential 
$-\frac{\partial}{\partial M} (F_R \rho_0 \Delta \x ) dM$ gives the total 
mass of collapsed halos in the mass range $M$ to $M + dM$. Hence, if 
$n_c(M,z) dM$ is defined as the spatial number density of halos between 
$M$ and $M+dM$ at $z$, we have
%eq3
\begin{equation}
-\frac{\partial}{\partial M} (F_R \rho_0 \Delta \x ) dM
=n_c(M,z) dM \cdot M\Delta \x,
\end{equation}
where we use the subscript $c$ in $n_c$ to emphasize that it is for 
collapsed halos. Considering the cloud-in-cloud problem, the above 
defined $n_c$ should be multiplied by a factor of 2. 
The normalization $\int_0^{\infty} n_c(M,z) M dM = \rho_0$ can then
be fulfilled. Therefore, the spatial number
density of halos with mass between $M$ and $M+dM$ at redshift $z$ is
given by
%eq4 
\begin{equation}
n_c (M, z)dM = -\frac{\rho_0}{M} \frac{\partial}{\partial M}
           {\rm erfc}\left(\frac{\delta_c(z)}{\sqrt{2}\sigma(R,z)}\right)dM,
\end{equation}
where erfc(x) is the complementary error function.

We will study models of SCDM, LCDM and CHDM, for which the density parameter 
$\Omega_0$, the cosmological constant $\lambda_0$, and $\sigma_8$, are listed 
in Table 1 (the Hubble constant is taken to be $H_0=100$ h km s$^{-1}$ Mpc$^{-1}$.) 
It is worth to point out that the $\sigma_8$ of the CHDM and LCDM models are 
compatible with the COBE-DMR observation, but that of the SCDM model is not. 
The value $\sigma_8=0.58$ gives a good fit for the SCDM model to all data 
except for the COBE result. We assumed that the initial primordial power 
spectrum is the Harrison-Zel'dovich type. The linear transfer functions of 
the SCDM and the LCDM models are taken from Bardeen et al. (1986) and that 
of the CHDM model from Klypin et al. (1995).

The cumulative number density, $N_c(M)$, of all halos with mass greater 
than 
$M$ should be
%eq5
\begin{equation}
N_c (M, z ) = \int ^{\infty} _M n_c (M_1, z ) dM_1.
\end{equation}
The abundance of halos calculated from Eq. (5) has been
verified by a number of N-body simulations (e.g. Lacey \& Cole 1994).

The redshift evolution of the number density of collapsed halos, $N_c(M, z)$,
are shown in Figs. 1a, 1b and 1c for the SCDM, LCDM and CHDM models, 
respectively. The eight curves in Figs. 1a and 1b correspond to 
$M=10^{13.5+n0.25} \ M_{\odot}$, with $n=0,1... 7$ from top to bottom. 
In Fig. 1c, it is $M=10^{11.5+n0.25} \ M_{\odot}$, with $n=0,1... 7$. 

Pairs and multiples of QSOs are not common to see. Therefore, it is 
reasonable to assume that each collapsed halo hosts only one, or at most 
a few QSOs. The number density of QSOs with magnitude $M_B < -26$
is plotted as crosses in Figure 1. The data points are taken
from Pei (1995) whose result is based on the observations of Hewett et al.
(1993) and Schmidt et al. (1992). In Pei's paper,
$n(z)$ is measured in the Einstein-de-Sitter cosmological model
($\Omega_0=1$, $\lambda_0=0$ and h = 0.5). When comparing the observations
to the LCDM model, we have made corrections of the cosmological effect on
$n(z)$ due to the non-zero $\lambda_0$.

 Fig. 1 shows that for all the models, the lower mass limits are mainly
determined by the QSO abundance at $z=4$. The possible hosts of QSOs in the
SCDM and LCDM models can be provided by halos with masses above
$10^{13.75}$, and $10^{14}\ M_{\odot}$, respectively. In the CHDM model, 
the number densities of collapsed halos at these mass range are much fewer. 
Therefore, for the CHDM, we have to include lower mass halos to host QSOs. 
The mass of possible QSO halos will be as low as about 
$M=10^{11.75} \ M_{\odot}$, which is more than two orders of magnitude 
below those in the models of SCDM and LCDM.

\section{Two-Point Correlation Functions of QSOs}

The spatial number density of uncollapsed spherical regions can also be
calculated in PS formalism. We define $m(M_0, r, z)dM_0dr$ to be the
number density of regions with mass $M_0 \to M_0 + dM_0$ and radius
$r \to r + dr$ at $z$. Similar to Eq.(4), we have
%eq6
\begin{equation}
m(M_0, r, z) =   -\frac{\rho_0}{M_0} \frac{1}{2} \frac{\partial ^2}
             {\partial M_0 \partial r}
    {\rm erfc}\left[\frac{\delta(r,z)}{\sqrt{2}\sigma(r_0,z)}\right].
\end{equation}
where $r_0$ is defined by $M_0 = V_0 \rho_0=4\pi r_0^3\rho_0/3$.
$\delta(r,z)$ is the overdensity, for which a region $M_0$ will evolve
into a sphere with radius $r$ at $z$. Function $\delta(r,z)$ is also
calculated in Bi \& Fang (1996). The total number of such spherical
regions in an arbitrary volume $\Delta \x$ should on average be given by
$m(M_0, r, z) dM_0 dr \Delta \x$.

The mass correlation function $\xi(r)$ is determined by the relative 
enhancement of mass density in the spherical shell $r \rightarrow r + dr$
around area $\Delta \x$. Only the spheres with radius
$r \rightarrow r + dr$ can contribute to this enhancement.
The mean enhancement of each $M_0$ sphere is approximately 
described by its mass variance $M_0^2\sigma ^2(r_0, z)$. Therefore, the mass
correlation function can be estimated by
%eq7
\begin{eqnarray}
\xi(r, z) & = & \frac{\int_0 ^{\infty} m(M_0,r,z) dM_0 dr \Delta \x
     M_0^2 \sigma ^2 (r_0, z)}
     {\rho_0 4\pi r^2dr \cdot \rho_0 \Delta \x} \\ \nonumber
     &  = &  \int_0^{\infty} dM_0 m(M_0, r, z) M_0 ^2 \sigma^2(r_0, z)/
     4\pi r^2\rho_0^2.
\end{eqnarray}
The factors $\rho_0 4\pi r^2dr$ and $\rho_0 \Delta \x$ are the mean mass in
the spherical shell $r \rightarrow r + dr$ and the volume $\Delta \x$,
respectively. In deriving Eq. (7), we implicitly assumed that there is
no correlation among the initial spheres, so the mass fluctuation in the
shell is simply given by the sum of the individual components.

\hbox to 8.5cm{
  \vbox to 6 cm{
  \vfil
\includegraphics{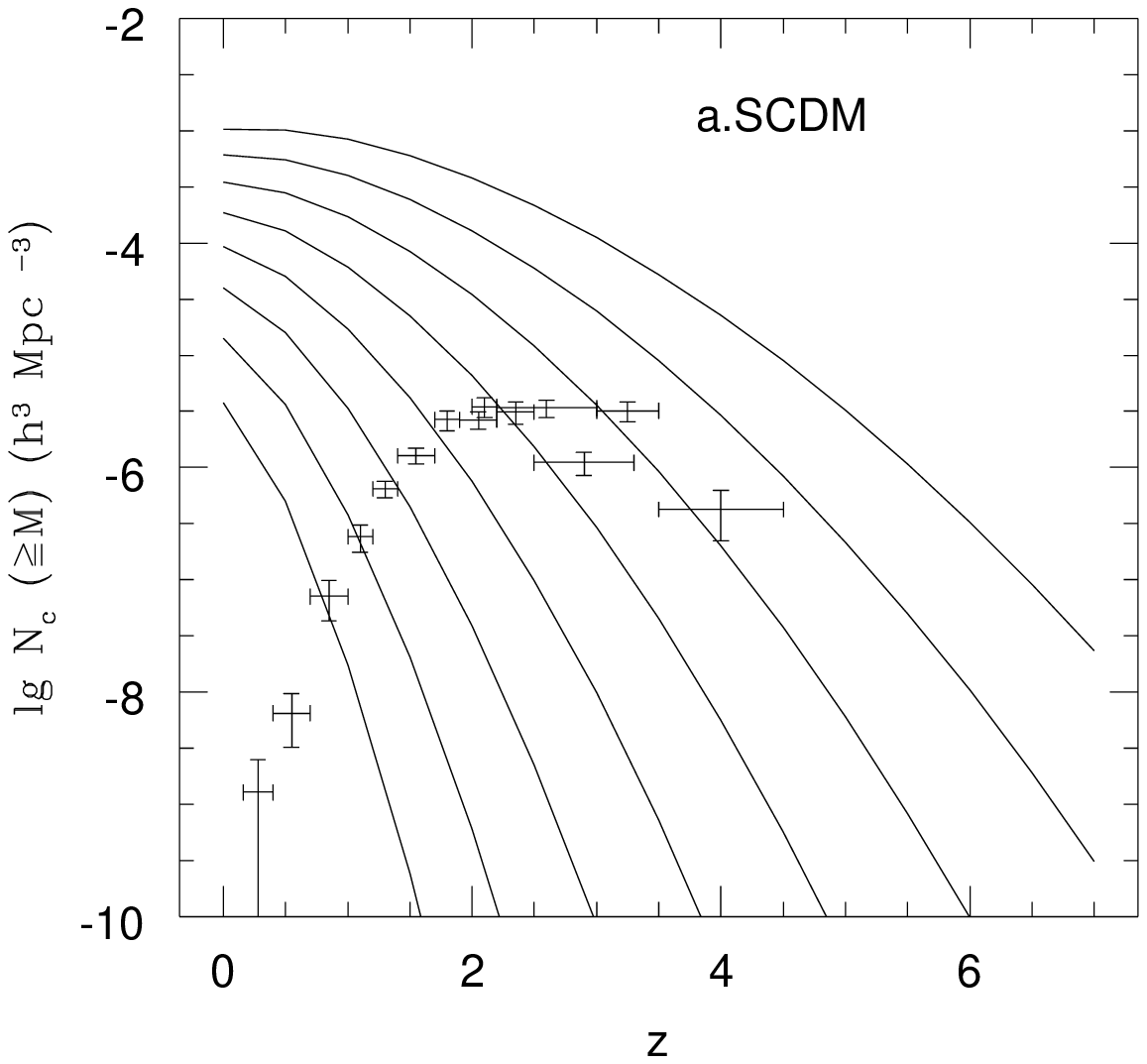}
}
%\centerline{
%}
}

\hbox to 8.5cm{
  \vbox to 6 cm{
  \vfil
\includegraphics{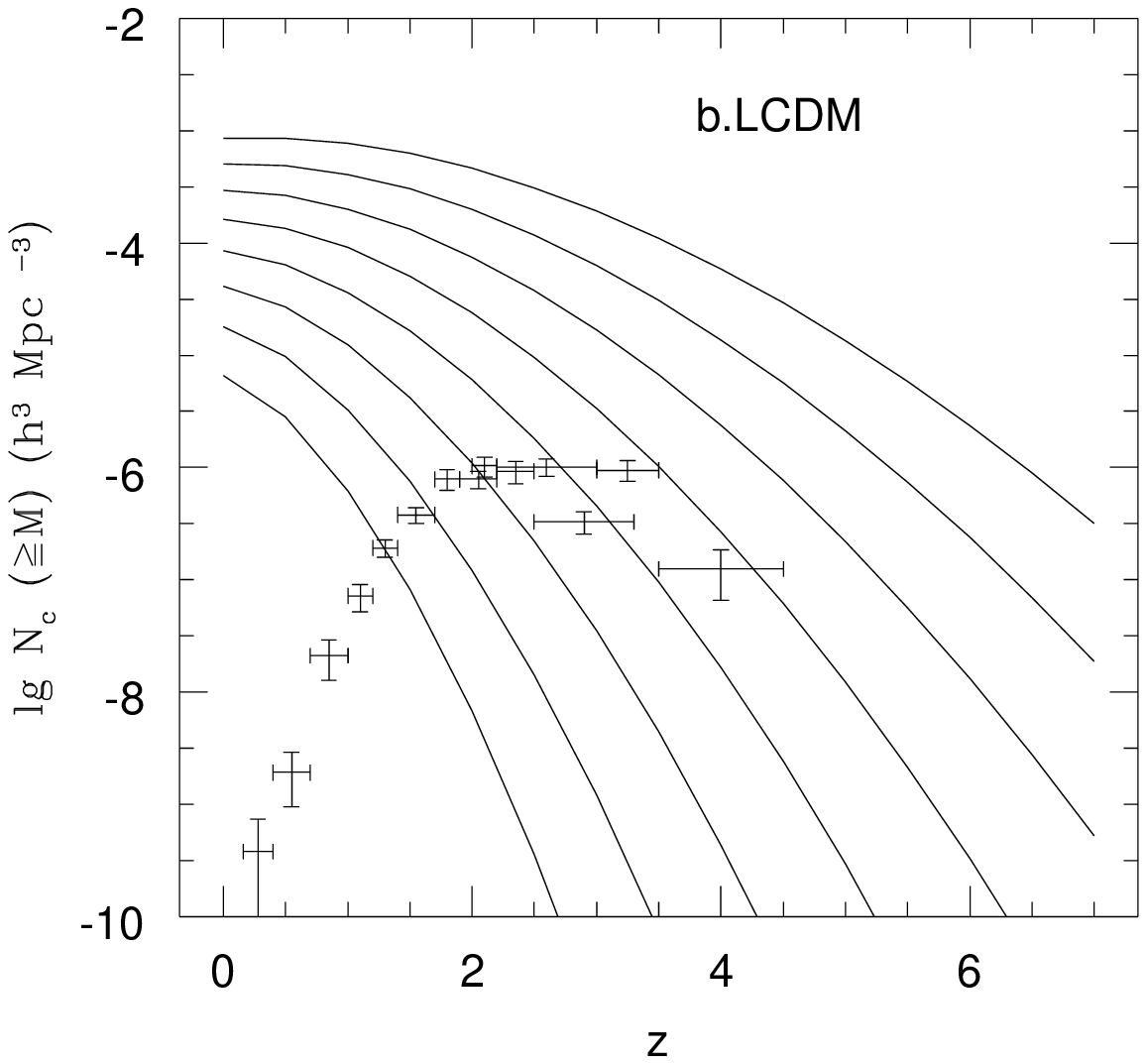}
}
%\centerline{
%}
}

\hbox to 8.5cm{
  \vbox to 6 cm{
  \vfil
\includegraphics{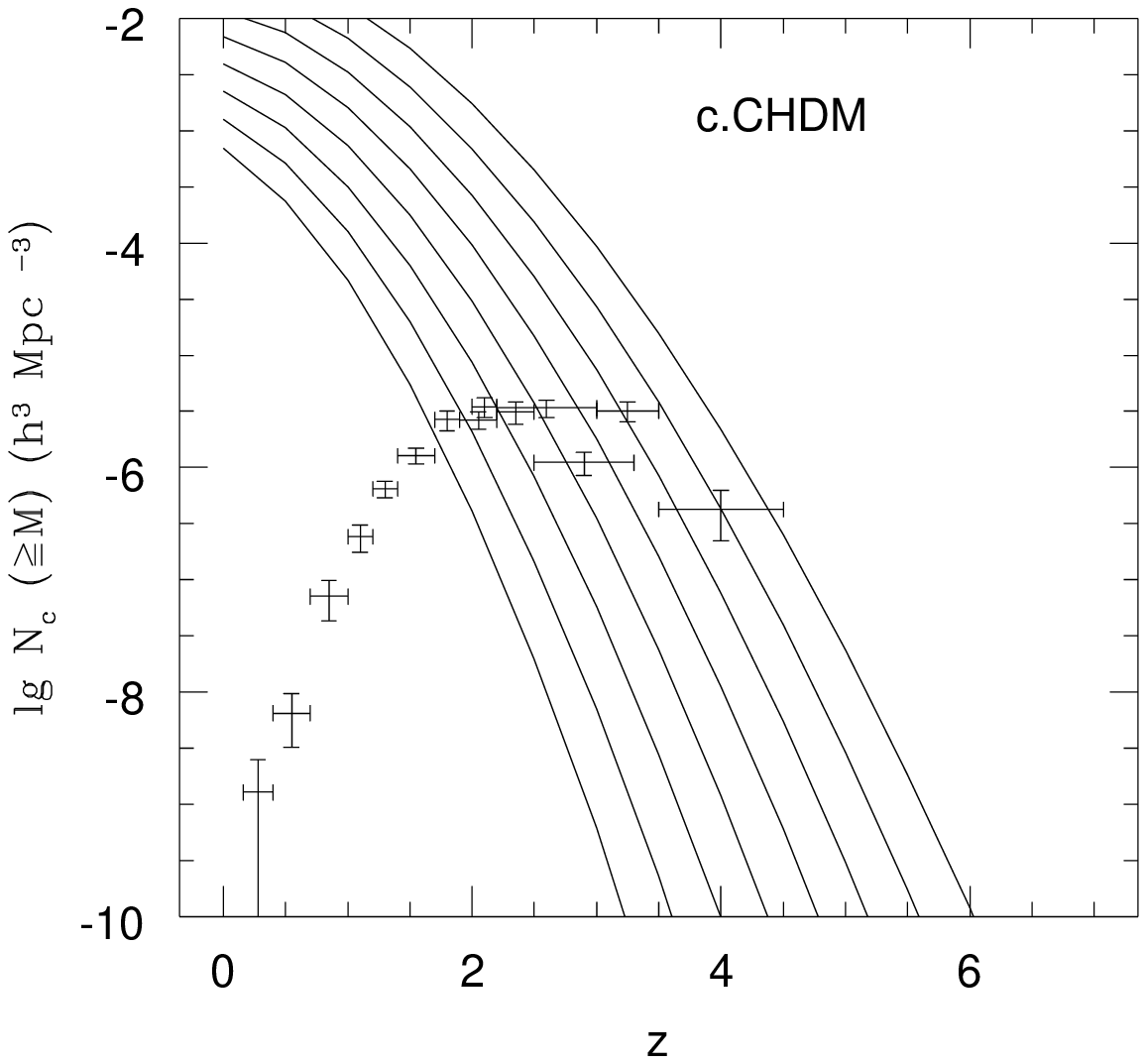}
}
%\centerline{
%}
}

\vspace{-2.2cm}
Fig. 1. Evolutions of the comoving number density of
halos with mass larger than a given number $M$.
a.) for SCDM model, the eight curves
correspond to $M=10^{13.5+n0.25} \ M_{\odot}$, with $n=0,1... 7$
from top to bottom. b.) The same as a.) but for LCDM model.
c.) for CHDM model, here $M=10^{11.5+n0.25} \ M_{\odot}$,
with $n=0,1... 7$.

\bigskip

Because it is impossible that a collapsed halo with mass $M$ formed in
a uncollapsed region with $M_0 < M$, the number density of $M$ collapsed
halos in an uncollapsed sphere of mass $M_0$ should be zero if
$M$ is greater than $M_0$. Thus, it is reasonable to assume that the number
density of collapsed halos in each PS sphere $r$ at $z$ on average is
%eq8
\begin{equation}
{\cal{N}} _c (M, M_0) = \left\{ \begin{array}{ll}
                        A N_c(M,z) & \mbox{for $M \le M_0$,} \\
                        0            & \mbox{for $M > M_0$,}
                        \end{array}
                     \right.
\end{equation}
where the constant $A \ge 1$ is introduced to maintain the normalization
condition
%eq9
\begin{equation}
\int ^{\infty} _0 {\cal{N}}_c V_0 n(M_0,z) dM_0 = N_c(M,z).
\end{equation}
We have then $A=2/ {\rm erfc} [\delta(R,z)/\sigma (R,z)]$.

Like calculating Eq.(7), let's consider a typical spherical shell
$r \rightarrow r+dr$. Only the spheres with radius
$r \rightarrow r + dr$ can contribute to the mass enhancement in
this shell. The total number of uncollapsed regions with radius
$r \rightarrow r + dr$ and masses $M_0 \rightarrow M_0 +dM_0$ in a
volume $\Delta \x$ is $m(M_0,r, z) dM_0 dr \Delta \x$. The total number
of collapsed halos $M$ in each $M_0$ sphere is ${\cal{N}}_c(M) V_0$, and
the variance of the number is $({\cal{N}} _c V_0)^2 \sigma ^2 (r_0,z)$.
Therefore, the correlation function of collapsed halos with mass larger
than $M$ can be  approximated as (Bi \& Fang 1996)
%eq10
\begin{eqnarray}
\lefteqn{\xi (r; \ge M, z) =} \nonumber \\
 & &  \frac{
\int_M^{\infty} dM_0 m(M_0, r,z)dr\Delta \x \cdot ({\cal{N}}_c V_0)^2
  \sigma^2(r_0,z)}{N_c4\pi r^2 dr\cdot N_c \Delta\x}  \nonumber \\
  & & A^2 \int_M ^{\infty} dM_0 m(M_0, r,z)V_0^2\sigma ^2 (r_0, z)/4\pi r^2
\end{eqnarray}

Eq.(10) implicitly assumed that in uncollapsed regions, the
collapsed halos have the same linear variance as the mass. The approximation
given by Eq.(10) is found to be in good agreement with the
linear approximation on scales larger than $R$, but higher than the
empirical formalism of Hamilton et al. (1991) on scales less than $R$,
here $R$ is the scale within which the non-linear effects are significant. 
Therefore, Eq.(10) can be used at least as a upper limit to the correlation
function.

Using Eq.(10), we have calculated the correlation function $\xi (r, \ge M)$ 
at redshift $z=2.5$. The results are shown in Fig. 2, in which 2a and 2b
are for the models of SCDM and LCDM, and 2c for CHDM. The eight curves in
Figs. 2a and 2b are corresponding to masses $10^{13.5+n0.25}M_{\odot}$,
$n=0, ...7$ from left to right, respectively. In Fig. 2c, it is
$10^{11.5+n0.25}M_{\odot}$ and $n=0, ...7$.

The two-point correlation function of QSOs is found to be obey the same power
law as galaxies $\xi_{qq}(r) = A_{qq} r^{-1.8} =(r/r_0)^{-1.8}$,
and the amplitude $A_{qq} \sim 25$ (or the correlation length
$r_0 \sim  6 \ h^{-1}$Mpc) at $z=1.5$ when $q_0$ is taken to be 0.5 
(Mo \& Fang 1993). This gives $\xi_{qq}(r) > 0.1$ for 
$r = $ 5 \ - \ 10 h$^{-1}$ Mpc. The clustering of QSOs on scales of 
$r\sim 30\ h^{-1}$Mpc is also found to be significant
(Deng et al. 1994). Some observations and statistics have even 
indicated the possible existence of groups of QSOs with comoving sizes as 
large as about $100\ h^{-1}$Mpc (e.g. Clowes \& Camppusano 1991,
Komberg et al. 1996.) Despite $A_{qq}$ is decreasing for $z\geq 1.5$, the
amplitude $\xi_{qq}$ is still larger than 0.1 (Mo \& Fang 1993;
Komberg et al. 1994.)

\vspace{2.2cm}

\hbox to 8.5cm{
  \vbox to 6.4cm{
  \vfil
\includegraphics{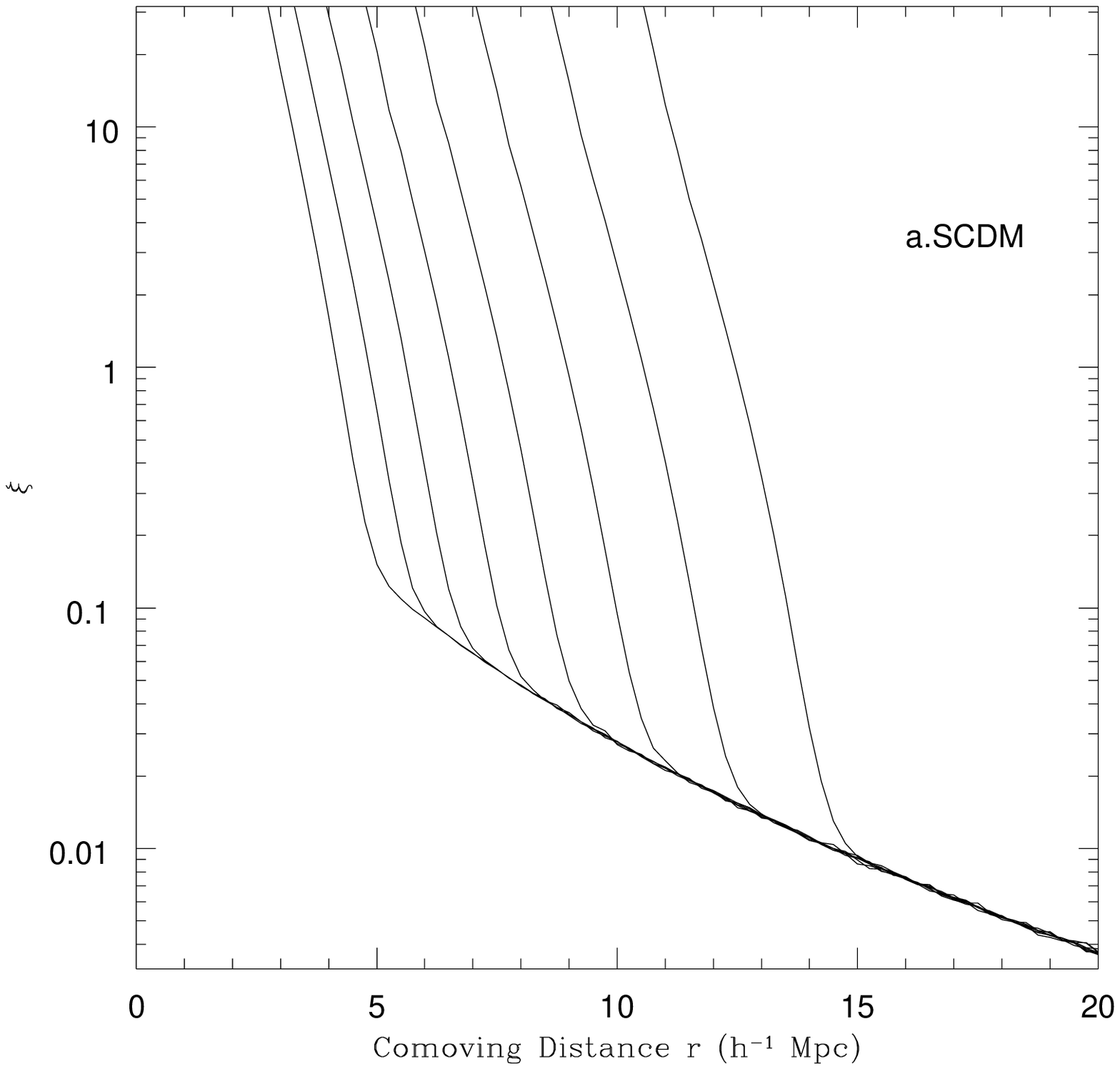}
}
%\centerline{
%}
}

\hbox to 8.5cm{
  \vbox to 6.4cm{
  \vfil
\includegraphics{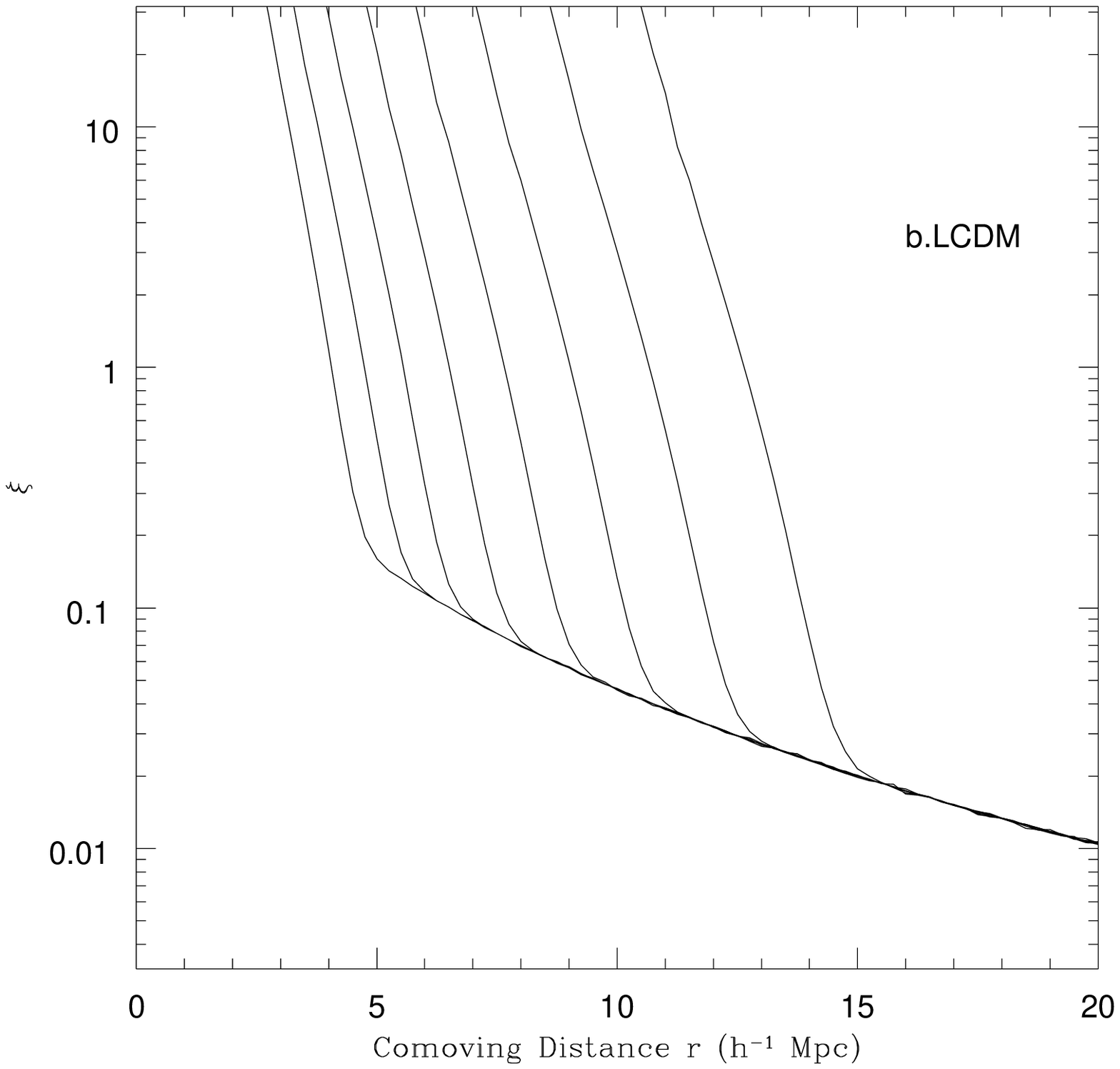}
}
%\centerline{
%}
}

\hbox to 8.5cm{
  \vbox to 6.4cm{
  \vfil
\includegraphics{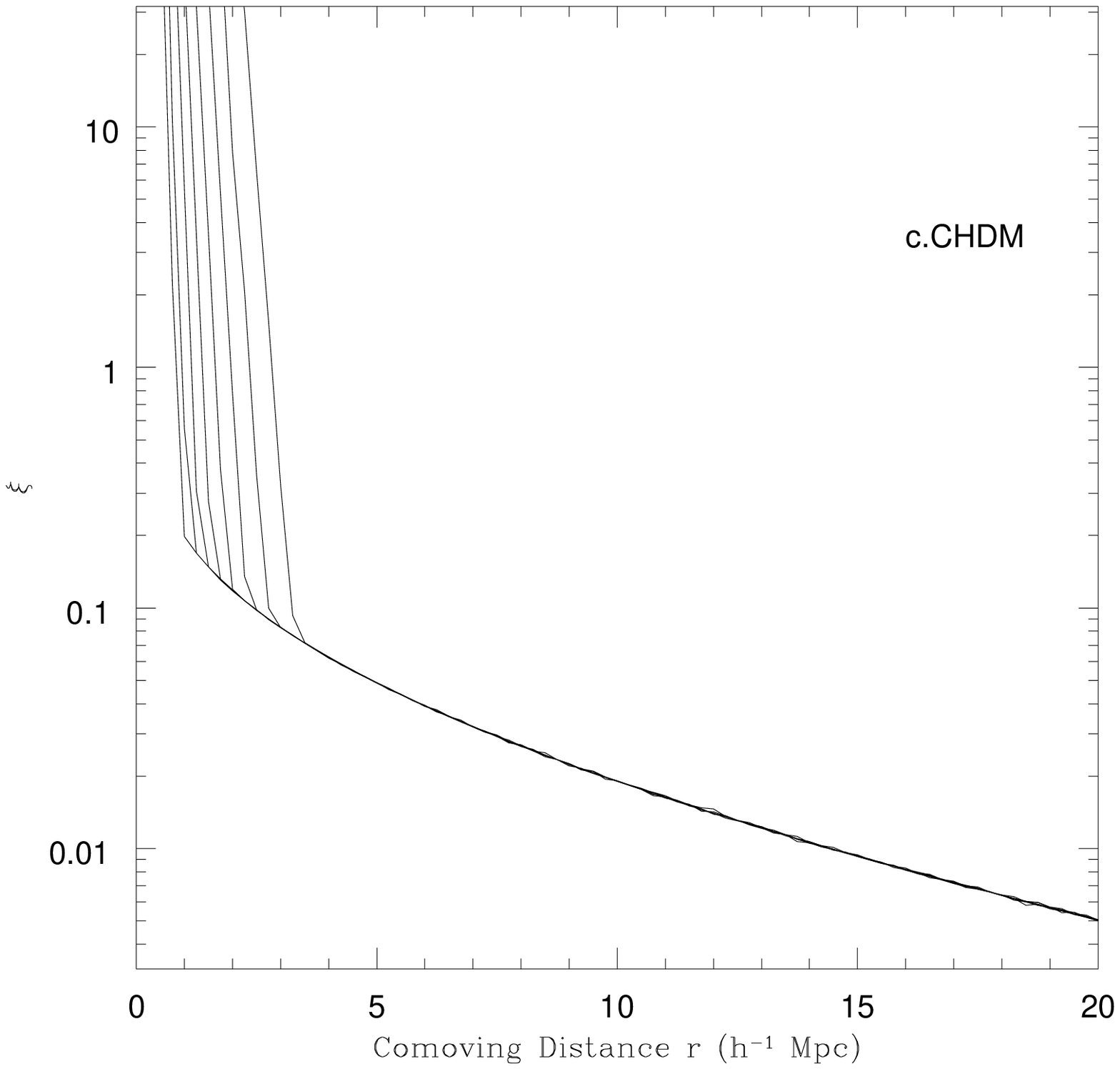}
}
%\centerline{
%}
}

\vspace{-2.4cm}
Fig. 2.  Two point correlation functions of $M$ halos
at $z=2.5$.  a.) For SCDM model, the eight curves from left to right
are corresponding to masses $M=10^{13.5+n0.25}M_{\odot}$ and $n=0, ...7$,
respectively. b.) The same as a.) but for LCDM model. c.) For
CHDM model, here $M=10^{11.5+n0.25}M_{\odot}$ and $n=0, ...7$.

\bigskip

Fig. 2a and 2b show that the correlation function of 
$M \geq 10^{14}M_{\odot}$ halos in the 
SCDM and LCDM models are consistent with the observational data. These halos
have also proper spatial number density as the observed QSOs.
The difference between the two models are very small. In fact, the top-hat
evolutions of a spherical mass in SCDM and LCDM are indistinguishable,
because it has almost the same dynamic trajectory in all flat 
universes with $\lambda_0 \leq 0.8$.

On the other hand, Fig. 2c shows that the two-point correlation functions
of the CHDM model are very small on large scales. For halos with mass  
$10^{13.25}M_{\odot}$, the amplitude of the correlation function is well less 
than 0.1 on scales of $r = $5 -10 h$^{-1}$ Mpc. Therefore, the CHDM model 
seems to be unfavored under the abundance-plus-correlation test. According to the
abundance, the mass of QSO hosts in the CHDM model should be as small as
$10^{12}M_{\odot}$. On the other hand, the two-point correlation function of
such halos is much less than what is observed.

\section{ Biasing problem}

A possible way to save the CHDM model is to assume that there
is a very high biasing factor to raise the correlation function of
$10^{11.5}M_{\odot}$ halos in the CHDM model. From Fig. 2c, if this
factor is as large as 5, the power of two-point correlation functions
mass $10^{11.5}M_{\odot}$ would be able to fit with QSO's clustering.
However, biasing is actually not a free parameter which can be
arbitrarily chosen (Einasto et al. 1994).

Generally speaking, the correlation functions of objects identified from
a mass field will be different from the mass itself.
A different method in the identification procedure can lead to a different 
biasing. The question now is: can we choose other QSO identifications
to produce larger biasing, and then to give higher amplitude of the
two-point correlation functions of QSOs in the model?

As an example, let us first consider the QSO identification by velocity
dispersion.
It has been known for a decade that QSOs with low redshift are preferentially
located in small groups of galaxies. This is in evidence from  QSO-galaxy
correlation function (Yee \& Green 1987), CIV-associated absorption in high
redshift radio-loud QSOs (Flotz et al. 1988), clustering analyses of QSO 
distribution (Bahcall \& Chokshi 1991) and the galaxy environments around 
QSOs (Ellingson et al. 1991a). It has been also shown that the velocity
dispersion of galaxies around QSOs is $ \sim$ 400  km s$^{-1}$ (Ellingson
et al. 1991b). Therefore the environment suitable of QSOs formation
 seems to be small groups of velocity dispersion 
$\rm \sim 400 \, km \, {\rm s}^{-1}$. The strength of the QSO correlation 
function, is intermediate between galaxies and rich clusters, so once again 
it is similar to galaxy groups.

If we assume that high redshift QSOs formed in the same environment as that
in the low redshift, the QSOs should be identified as collapsed
halos with 3-dimensional velocity dispersion
$\sigma_v \sim \sqrt{3} \times 400 = $ 700 km s$^{-1}$
between $z=2$ and $z=4$.

In the PS formalism, the comoving number density of halos with velocity
dispersion $\sigma_v$ can be calculated by
%eq11
\begin{eqnarray}
\lefteqn{n(\sigma_v,z) d\sigma_v = 
   -{3 \over (2\pi)^{3/2} R^3 } } \nonumber \\
 & & \times {d\ln \sigma(R,z)\over d \ln R} {\delta(z) \over \sigma(R,z)}
 \exp \left(-{\delta_c^2(z) \over \sigma^2(R,z)} \right)
{dR\over R}\,
\end{eqnarray}
As in Eq.(4), $\delta_c(z)$ in Eq.(11) is the critical overdensity for 
collapsed at
redshift $z$. The meaning of $R$ here is the same as in \S 2.
 The relationship between $\sigma_v$ and $R$ is given by (Narayan 
\& White 1988)
%eq12
\begin{equation}
\sigma_v= c_\sigma \sqrt{3} H_0 R (1+z)^{1/2}
\end{equation}
for Einstein-de-Sitter universe, and
%eq13
\begin{equation}
\sigma_v= c_\sigma \sqrt{3} \Omega_0^{1/2} H_0 R
(1+z)^{1/2} .
\end{equation}
for open universe or the LCDM universe. The coefficient $c_{\sigma}$ 
has been determined by comparing
Eqs. (11) and (12) with N-body simulation (Jing \& Fang 1994). It found
$c_\sigma=1.1 \ - 1.3$. For most calculations,
$c_\sigma=1.2$ is a preferred value. Hence, one can safely use Eqs.(11)-
(13) even when the 3-D velocity dispersion is as large as about
800 km s$^{-1}$.

The total number density of the collapsed halos with the
velocity dispersion greater than a certain value, say $\sigma^{lim}$, is
%eq14
\begin{equation}
N(>\sigma^{lim}, z)=\int^{\infty}_{\sigma^{lim}}
n(\sigma_v,z)d\sigma_v.
\end{equation}

Using Eq.(14), we found that the SCDM and LCDM models can produce sufficient
number of $\sigma^{lim} =$ 700 km s$^{-1}$ halos to fit with the number
density of QSOs. But the number of such halos in the CHDM model is too small.
To have enough number of halos, we should use $\sigma^{lim} = $ 300 km s$^{-1}$
or less in the CHDM model.

The mass of the $\sigma_v$-selected halos can be determined from Eqs.(12)
or (13). One can then calculate the two-point correlation functions
of them by Eq.(10). Here we get almost the same results as before
because $\sigma_v$ is one-to-one related to the mass or the radius of
the top-hat windows. Similar to Fig. 2c, the $\sigma^{lim} =$ 300 km s$^{-1}$ 
halos still lack of correlation power on all scales larger than 2 h$^{-1}$ 
Mpc in the CHDM model.

This result should be expected. An identification of
objects from a density field is a sampling. As one knows, a sampled
field will not be different from the original field on scales much larger 
than the characteristic scale, $R_c$, of the sampling (Vanmarke 1983).
This means that, in principally, no biasing on scales larger than the
characteristic scale can be introduced by the identification. For our 
question here, it is not easy to given the characteristic scale $R_c$ of 
the identification. However, it is probably reasonable to choose 
$R_c= [1+\overline{\xi(R,z)}]^{1/3}R$, where $\overline{\xi(R,z)}$ is the 
average of 
$\xi(r,z)$ inside $R$ (Hamilton et al. 1991). So the scale $R_c$ is actually 
the coherence length of the density field smoothed by window $R$. The 
distribution of halos should not be biased from mass distribution on scales 
larger than $R_c$. The correlation function can effectively be amplified by
linear geometrical biasing only for large halos (Kaiser, 1984).
Figure 2 show a break in the correlation functions at $r_b$, which is roughly
equal to $R_c$. When $r > r_b$, the correlation functions of the
collapsed halos should approach to the mass correlations, while when
$r<r_b$, the biasing leads to larger correlation functions.

All identifications based on gravitational parameters,  such as
circular velocity, virial temperature etc. are essentially equal to mass 
identifications, because for a gravitational collapsed systems these
parameters are one-to-one related. Therefore, they can be expressed 
as an equivalence to $M$. Biases in these parameters cannot give significant 
different results from those using $\sigma_v$ or $M$.

Obviously, any non-gravitational identifications are not constrained by
the gravitational characteristic scale. For instance, if gas processes in
the formation of QSOs play the role of biasing, the characteristic scale
will not equal to $R_c$. However, the characteristic scale of gas processes
should be much less than $R_c$, because the velocity of gaseous component is
too small to segregate the QSO halos on scales of 5 h$^{-1}$ Mpc.

\section{ Conclusions}

We showed that the abundance-plus-correlation test can serve as a
promising discrimination among models of cosmic structure formation using
the high redshift ($z>2$) QSO number density and their clustering on large
scales (10 h$^{-1}$ Mpc). We found that the SCDM and LCDM model are
consistent with the observations of abundance and two-point correlation,
while the CHDM models is difficult to fit with the observed numbers.

The contributions of variously biasing mechanisms have been investigated.
A bias, in fact, is a model of the environment suitable to form objects,
or a phenomenological relationship between the cosmic density field and QSOs. 
We have studied possible biasing processes, including both gravitational
and non-gravitational origins. None of them seems to give large
characteristic scale and clustering amplitude required to make the
CHDM model success.

One should also consider the possibility that each halo may host more than
one QSOs. In this case, the correlation function can be stronger than that
of one-QSO/one-halo model. In order to fit with observed abundance of QSOs
in the CHDM model, one can assumed that each halo of $M=10^{14}M_{\odot}$ 
host more than 10 QSOs. In this case, halos of $M=10^{14}M_{\odot}$ in the 
CHDM can also fit with the observed two-point correlation function of QSOs 
(Fig. 2c). However, this apparently leads to the paradox that we did not 
found so many bright QSOs concentrating in galaxy groups. A possible way
to explain it is that the lifetime of QSOs is just longer than the cosmic
time at $z=3$ but shorter than the cosmic time at the present. (If QSOs have
a life time shorter than the cosmic time at $z=3$, we require even more low
mass halos, this makes the situation even worse.) However, the difference
between the cosmic time at $z=3$ and $z=0$ is only a factor of 8
($t\propto (1+z)^{3/2}$ in the $\Omega=1$ universe). Therefore, we may predict
that on average, each of 8 galaxy group will have a bright QSO. This is not
true. Otherwise, we should require a very particular mechanism to make the
QSO formation at $z=3$.

It is theoretically possible to explain any correlations if we
are allowed to introduce {\it unknown} inhomogeneity into the density field,
and assume that the correlation of QSOs absorbers is given by these
inhomogeneities. However, to plan these inhomogeneities is equal to put 
desired structures in the initial perturbations. Such models will, however, 
no longer be the SCDM, LCDM or CHDM models, which are based on initial 
fluctuations produced at the inflationary era. However, even in this case 
the CHDM will still be difficult to produce enough halos with reasonable
velocity dispersions to fit with the abundance of high redshift
QSOs.

\acknowledgements
We benefit by the discussions and comments from Yipeng Jing and Houjun Mo, 
and thank J. Einasto for suggestions given in his referee report. HGB
thanks also for support of World Lab during the first phase of this work.

\end{document}